\documentclass[doublecol,linenumbers]{epl2} 

\usepackage{subfigure}
\usepackage{graphicx}
\usepackage{amssymb}
\usepackage{amsmath}
\usepackage{bbm}
\usepackage{placeins}
\DeclareMathAlphabet{\cali}{OMS}{zplm}{m}{n}

\newcommand\ii{\mathrm{i}}
\newcommand\dd{\delta\:\!}

\title{Effect of the next-nearest-neighbor hopping on the charge collective 
modes in the paramagnetic phase of the Hubbard model}
\shorttitle{Effect of the NNN hopping on the charge collective modes in the 
paramagnetic phase of the Hubbard model} 

\author{Vu Hung Dao\inst{1} \and Raymond Fr\'esard\inst{1}}
\shortauthor{V. H. Dao \etal}

\institute{                    
  \inst{1} Normandie Univ, ENSICAEN, UNICAEN, CNRS, CRISMAT, 14000 Caen, France
}

\abstract{
The charge dynamical response function of the $t-t'-U$ Hubbard model is 
investigated on the square lattice in the thermodynamical limit. 
The correlation function is calculated from Gaussian fluctuations around the 
paramagnetic saddle-point within the Kotliar and Ruckenstein slave-boson 
representation. 
The next-nearest-neighbor hopping only slightly affects the renormalization of
the quasiparticle mass. In contrast a negative $t'/t$ notably decreases (increases) 
their velocity, and hence the zero-sound velocity, at positive (negative) doping.
For low (high) density $n \lesssim 0.5$ ($n \gtrsim 1.5$) we find that
it enhances (reduces) the damping of the zero-sound mode. Furthermore it
softens (hardens) the upper-Hubbard-band collective mode at positive (negative) 
doping. 
It is also shown that our results differ markedly from the random phase
approximation in the strong-coupling limit, even at high doping, while they compare 
favorably with existing quantum Monte Carlo numerical simulations.}

\begin{document}

\maketitle

\section{Introduction}
\label{sec:int}

Spin and charge excitation spectra of correlated fermionic systems may
be conveniently accessed using Kotliar and Ruckenstein (KR) slave-boson
representations of the microscopic model of interest
\cite{Li89,Rasul88,Lav90,li91,Kot92,FW,li94,Zim97,Dao17}. 
For instance, in the case of the Hubbard model on the square lattice, 
it has been recently shown that charge excitation spectra generically 
consist of a low-energy continuum, a zero-sound (ZS) collective mode, and 
another collective mode dispersing at energies scaling with the interaction 
strength~\cite{Dao17}. Hence these excitation spectra display the physics
contained in the concepts introduced by Landau in his theory of the
Fermi Liquid~\cite{Lan56}, and by Hubbard who established the splitting
of the band due to the Coulomb interaction~\cite{Hub63} which can now be
incorporated in a single calculation. While earlier attempts suffered
from various drawbacks~\cite{Li89,Rasul88,Lav90,li91,
Kot92,FW,li94,Zim97} the one-loop calculation of spin and charge 
susceptibilities was recently shown to comply with lowest order perturbation 
theory and particle-hole symmetry~\cite{Dao17,Dao17b}. 
What happens when the latter is broken? 
Hubbard-type models are thought to contain the low-energy physics of 
superconducting cuprates, the phase diagrams of which are not symmetric under
a sign change of the doping. One way to break the particle-hole symmetry is to 
take into account the next-nearest-neighbor (NNN) hopping. The purpose of this 
work is precisely to establish the influence of the latter on the charge excitation 
spectra, with parameter values generally accepted for the cuprates. 
We focus on the paramagnetic phase, free of symmetry breaking, in the 
thermodynamical limit  and we resolve the full momentum dependence of 
the spectra. Owing to their weak temperature dependence, and to the 
relatively low magnetic-instability temperature ($T_{\rm inst} \approx t/6$), 
our results essentially apply to the entire phase diagram.
 
We perform our investigations within the KR
slave-boson representation, which is able to capture interaction effects 
beyond the physics of Slater determinants. This approach reproduces 
the Gutzwiller approximation on the saddle-point level~\cite{Kot86}, 
which harbors the interaction driven Brinkman-Rice metal-to-insulator 
transition~\cite{Bri70}. Many valuable results have been obtained with 
KR~\cite{Kot86} and related slave-boson representations~\cite{Li89,FW}. 
For example the anti-ferromagnetic~\cite{Lil90}, spiral~\cite{Fre91,Fre92,Igo13,Doll2}, 
and striped~\cite{Sei01,Fle01,Sei02,Rac06,RaEPL} phases have been described 
with these methods, as well as the competition between the latter two~\cite{RaEPL}. 
Furthermore, it has been shown that the 
spiral order continuously evolves to the ferromagnetic order in the
large $U$ regime ($U \gtrsim 60t$)~\cite{Doll2}. Consistently, in the two-band model, 
ferromagnetic instabilities were found in the doped Mott insulating 
regime only~\cite{Fre02}. Yet, ferromagnetic instability lines arise in the 
intermediate-coupling regime either through the introduction of a ferromagnetic
exchange coupling~\cite{lhoutellier15}, or due to a sufficiently large NNN hopping 
amplitude~\cite{FW98}, or on the fcc lattice~\cite{Igo15}.
The framework has been used most recently to address strong correlation effects
in the plates of a capacitor and a possible capacitance gain~\cite{Ste17}. 
Furthermore, the comparison of groundstate energies to existing numerical simulations 
on the square lattice showed that the difference between the numerical estimate 
and the slave-boson result is less than 3\% for $U=4t$~\cite{Fre91}.
For larger values of $U$ and doping larger than 15\%, it has been obtained 
that the slave-boson groundstate energy exceeds the exact diagonalization 
data by less than 4\% for $U=8t$, and less than 7\% for $U=20t$. 
The discrepancy increases when the doping is lowered~\cite{Fre92}.
In addition, quantitative agreement to quantum Monte 
Carlo (QMC) charge structure factors was established~\cite{Zim97}.
Furthermore, it has been shown in \cite{Dao17} that time-dependent
Gutzwiller approximation and slave boson calculations at the Gaussian level
exhibit both qualitative and quantitative differences, despite the
exact correspondence between the Gutzwiller approximation and the
saddle-point approximation to the KR representation.

The letter is organized as follows. Firstly we give a 
brief presentation of the spin-rotation-invariant (SRI) KR slave-boson 
representation of the Hubbard model and the method 
used to calculate dynamical response functions (more details can be found 
in, e.g., review~\cite{fresard12}). Then we evaluate the 
charge susceptibility from fluctuations captured within the one-loop 
approximation, and investigate the dispersion of its
collective modes. Lastly we summarize the letter in the conclusion. 

\section{Model and method} \label{sec:method}

Within the SRI KR slave-boson representation~\cite{Kot86,Li89,FW,fresard12}
the Hubbard Hamiltonian is expressed as
\begin{equation} \label{eq:model}
 H = \sum_{i,j} t_{ij}\sum_{\sigma, \sigma', \sigma''} z_{i\sigma'' \sigma}^{\dagger} 
 f_{i\sigma}^{\dagger} f_{j\sigma'} z_{j\sigma' \sigma''} + U \sum_i d_i^{\dagger} d_i
\end{equation}
 with auxiliary-boson operators $e_i$, $p_{i\mu}$, $d_i$ (for atomic states with 
 respectively zero, single and double occupancy) and pseudo-fermion operators $f_{i\sigma}$. 
Note that in the approach, the on-site Coulomb interaction is represented by a term 
bilinear in bosonic operators. 
Yet this is at the expense of the hopping term, which 
is supplemented by the occupancy-change operator $z_{i\sigma\sigma'}$.
 In order to preserve spin rotation symmetry the latter is
defined as
\begin{equation}
 {\underline z}_i = e_i^{\dagger} {\underline L}_i M_i {\underline R}_i \, 
 {\underline p}_i + {\underline {\tilde{p}}}_i^{\dagger} {\underline R}_i M_i  
 {\underline L}_i \, d_i
\end{equation}
with 
\begin{eqnarray}
& M_i & = \Big[ 1 + e_i^{\dagger} e_i + \sum_{\mu=0}^3 p_{i\mu}^{\dagger} p_{i\mu} 
+  d_i^{\dagger} d_i \Big]^{1/2}, \nonumber \\
&{\underline L}_i & = \Big[ (1 -d_i^{\dagger} d_i) {\underline \tau}^0 
- 2 {\underline p}_i^{\dagger} {\underline p}_i \Big]^{-1/2}, \nonumber \\  
& {\underline R}_i & = \Big[ (1 - e_i^{\dagger} e_i) {\underline \tau}^0 
- 2 {\tilde{\underline p}}_i^{\dagger} {\tilde{\underline p}}_i \Big]^{-1/2},          
\end{eqnarray} 
and the $2\times2$ matrices in spin space 
${\underline p}_i = \frac{1}{2} \sum_{\mu=0}^3 p_{i\mu} {\underline \tau}^{\mu}$ 
and $\tilde{\underline p}_i = \frac{1}{2} ( p_{i0} {\underline \tau}^0 
- {\bf p}_i \cdot \boldsymbol{\underline \tau} )$, which are built from
the canonical operators $p_{i\mu}$, the identity matrix $\underline{\tau}^0$, 
and the Pauli matrices (for more details, see \cite{fresard12}).

Furthermore the auxiliary-boson operators generate a Fock space that contains 
more states than the physical ones. By definition the latter have exactly one atomic 
state per site, which means they belong to the subspace where, on each site $i$, 
the operator equality 
 \begin{equation}
 e_i^{\dagger} e_i + \sum_{\mu=0}^3 p_{i\mu}^{\dagger} p_{i\mu} 
 +  d_i^{\dagger} d_i = 1
\end{equation}
is satisfied. They additionally comply with the constraints 
\begin{subequations}
 \begin{align}
  \sum_{\mu=0}^3 p_{i\mu}^{\dagger} p_{i\mu} + 2 d_i^{\dagger} d_i 
  & = \sum_{\sigma} f_{i \sigma}^{\dagger} f_{i \sigma}, \label{eq:const2a} \\
  p_{i0}^{\dagger} {\bf p}_i + {\bf p}_i^{\,\dagger} p_{i0} 
  - \ii {\bf p}_i^{\,\dagger} \times {\bf p}_i & = \sum_{\sigma, \sigma'} 
  \boldsymbol{ \tau}_{\sigma \sigma'} f_{i\sigma'}^{\dagger} f_{i\sigma}, \label{eq:const2b}
 \end{align}
\end{subequations}
which equate the number of fermions to the number of $p$ and $d$ bosons. 
When calculating the partition function as a functional integral~\cite{li91,Zim97}, 
the physical constraints are then enforced with Lagrange multipliers $\alpha_i$ and 
$\beta_{i\mu}$. The internal gauge symmetry of the representation allows to 
gauge away the phases of $e_i$ and $p_{i\mu}$ by 
promoting the Lagrange multipliers to time-dependent fields~\cite{FW}, leaving us 
with radial slave-boson fields~\cite{Fre01}. Their saddle-point values may be viewed 
as an approximation to their exact expectation values that are generically 
non-vanishing~\cite{Kop07}. However, the slave-boson field corresponding to double 
occupancy $d_i = d'_i + {\rm i} d''_i$ has to remain complex~\cite{Jol91,Kot92,FW}. 

\begin{figure}
  \includegraphics[trim=0cm 0cm 0cm 0cm, clip=true, width=0.49\textwidth]
  {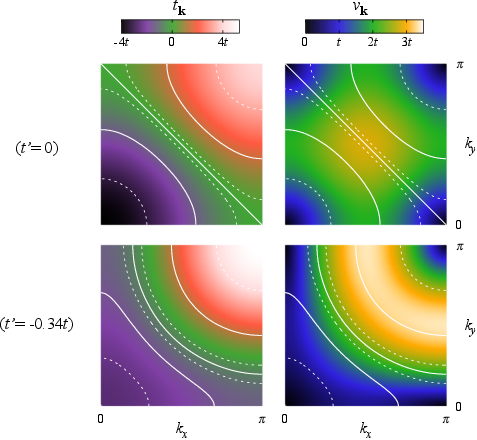}
  \caption{(Color online) Bare quasiparticle energy $t_{\bf k}$ and velocity 
  $v_{\bf k}$  on the square lattice for $t' =0$ and $-0.34t$. 
  The white lines are the Fermi surface plotted for density values $n=0.125$, 
  0.5, 0.875, 1, 1.125, 1.5, and 1.875 (from the lower left to the upper right 
  corner of each map).}
  \label{fig:dispersion}
\end{figure}

Within the saddle-point approximation, the quasiparticle mass is divided by
a factor $z_0^{2}$, which also plays the role of a quasiparticle residue.
For the paramagnetic solution 
 \begin{equation}
 z_0^2 ={\frac{2 p_0^2(e+d)^2}{1-\delta^2}},
 \end{equation}
where $e$, $p_0$, and $d$ are the saddle-point values of the boson fields, and 
$\delta = 1 - n$ is the hole doping from half-filling. The quasiparticle 
dispersion is renormalized as 
\begin{equation}
 E_{{\bf k}} = z_0^2 t_{{\bf k}} - (\mu - \beta_0)
\end{equation}
with $\mu$ the chemical potential, and $\beta_0$ the saddle-point value of
the Lagrange multiplier enforcing the constraint~(\ref{eq:const2a}). The bare 
quasiparticle energy is
\begin{equation}
t_{\bf k} = -2 t (\cos k_x + \cos k_y) - 4 t' \cos k_x \cos k_y 
\end{equation}
on the square lattice, when hopping processes of amplitude $t_{ij}=-t$ between 
nearest-neighbor sites and $t_{ij}=-t'$ between NNN ones are taken into account. 
In the absence of the latter, the energy and the 
velocity $v_{\bf k} = |\partial t_{\bf k}/\partial {\bf k}|$ are symmetric with 
respect to the Fermi level at half-filling $n=1$ (see the upper panel of 
Fig.~\ref{fig:dispersion}). The quasi-parabolic dispersion in the vicinity 
of the ${\bf k}$-points $\Gamma = (0,0)$ and ${\rm M}=(\pi,\pi)$ results in a 
nearly circular Fermi surface at large doping $|1-n|\approx 1$.
The lower panel of Fig.~\ref{fig:dispersion} shows that adding a finite hopping 
between NNN breaks the doping-reversal symmetry of the dispersion. 
A negative $t'$ (for $t>0$) non-uniformly increases the energy and enhances the 
velocity, as well as the isotropy of the dispersion, around M at the expense of 
the vicinity of $\Gamma$. Hence the Fermi velocity is quasi-isotropic at the
density $n>1$ for $t'=-0.34t$.

The saddle-point approximation is exact in the large degeneracy limit, 
while the Gaussian fluctuations are of order $1/N$~\cite{FW}. 
In addition it obeys a variational principle in the limit of large spatial 
dimensions where the Gutzwiller approximation becomes exact for the 
Gutzwiller wave function~\cite{Met89}. 
Within the Gaussian fluctuation approximation, the action is 
expanded to second order in field fluctuations
\begin{eqnarray}
&\psi(k) = \big(\delta e(k),\delta d'(k),\delta d''(k),\delta p_{0}(k),
\delta \beta_{0}(k), \delta \alpha(k), \nonumber \\ 
& \delta p_{1}(k),\delta \beta_{1}(k),\delta p_{2}(k),\delta \beta_{2}(k),
\delta p_{3}(k), \delta \beta_{3}(k) \big)
\end{eqnarray}

around the paramagnetic saddle-point solution 
$ \psi_{\rm MF} = (e, d, 0, p_0,\beta_0,\alpha, 0,0,$ $0,0,0,0) $
as
\begin{equation}
 \int d\tau {\cal L}(\tau) = {\cal S}_{\rm MF} + \sum_{k,\mu,\nu} \psi_{\mu}(-k) 
 S_{\mu\nu}(k) \psi_{\nu}(k)
\end{equation}
(the matrix $S$ is given in Ref.~\cite{Dao17}).
Here $k=({\bf k},\nu_n)$ with the bosonic Matsubara frequency $\nu_n=2\pi nT$, and 
$\sum_k = T \sum_{\nu_n} L^{-1} \sum_{{\bf k}}$ with $L$ the number of 
lattice sites. The correlation functions of boson fields are then 
Gaussian integrals which can be obtained from the inverse of the fluctuation 
matrix $S$ as 
$\langle \psi_{\mu}(-k) \psi_{\nu}(k) \rangle = \frac{1}{2} S^{-1}_{\mu\nu}(k)$. 
Using the density fluctuation 
$\delta {\cali N} = \delta(d^{\dagger} d - e^{\dagger} e) $, 
the charge susceptibility is
\begin{align}
 \chi_c(k) & = \langle \dd {\cali N}(-k) \dd {\cali N}(k) \rangle \nonumber \\
 &  = 2 e^2 S_{1,1}^{-1}(k) - 4 e d S_{1,2}^{-1}(k) + 2 d^2  S_{2,2}^{-1}(k) .
 \label{eq:chi_c-def}
\end{align}
The dynamical response function is eventually obtained within the analytical
continuation $\ii \nu_n \rightarrow \omega + \ii 0^+$. 
In the present work, for temperature $T=t/100$, the infinitesimal 
imaginary part is set between $10^{-6}t$ and $10^{-4}t$, depending on the needed 
frequency resolution.

The evaluation of the correlation functions in the paramagnetic 
state~\cite{Lav90,Zim97,li91,li94} yields the charge dynamical response 
function~\cite{Dao17}
\begin{equation}\label{eq:chi_c}
 \chi_c(k) = \frac{A(k) + B(k)  (\omega + \ii 0^+)^2}{C(k) + D(k)  
 (\omega + \ii 0^+)^2}
\end{equation}
where
\begin{align}\label{eq:coeff}
 A(k) = &\;  \tilde{S}_{33} \big[ 2 p_0^2 \Gamma_1(k) - 8 d p_0 \Gamma_2(k) 
 + 8 d^2 \Gamma_3(k) \big] , \\
 B(k) =  &\; 2 e d p_0^2 S_{55}(k) , \nonumber \\
 C(k) =  &\;  \tilde{S}_{33} \big[\Gamma_2^2(k) - \Gamma_1(k) \Gamma_3(k) \big]
 / e^2 S_{55}(k) , \nonumber \\
 D(k) = &\,  - \frac{ d \big[ p_0^{2\,}  \Gamma_1(k) + 2 (e-d)p_0 \Gamma_2(k) 
 + (e-d)^{\! 2 \,} \Gamma_3(k) \big]}{e (e+d)^2} , \nonumber
\end{align}
with
\begin{align}
 \Gamma_1(k) \, = &\; -S_{55}(k) [ e^2 S_{22}(k) - 2 e d S_{12}(k) + d^2 S_{11}(k) ] 
  \nonumber \\
 &\; + [e S_{25}(k) - d S_{15}(k)]^2, \nonumber \\
 \Gamma_2(k) \, = &\; -S_{55}(k) [ e^2 S_{24}(k) - p_0 e S_{12}(k) - e d S_{14}(k) 
 \nonumber \\
 &\; + d p_0 S_{11}(k) ] + [e S_{25}(k) - d S_{15}(k)] \nonumber \\
 &\; [e S_{45}(k) - p_0 S_{15}(k)], \nonumber \\
 \Gamma_3(k) \, = &\; -S_{55}(k) [ e^2 S_{44}(k) - 2 e p_0 S_{14}(k) + p_0^2 S_{11}(k) ] 
 \nonumber \\
 & \;+ [e S_{45}(k) - p_0 S_{15}(k)]^2 , \nonumber \\
 \tilde{S}_{33} \, = &\; -\frac{ 2 p_0^2 }{1 - \delta^2}  \varepsilon_0. 
\end{align}
Here the semi-renormalized kinetic energy is 
 \begin{equation}
 \varepsilon_0 =  \frac{2}{L} \sum_{{\bf k}}  t_{{\bf k}} n_F(E_{{\bf k}})
 \end{equation}
with the Fermi function $n_F(\epsilon) = 1/(\exp(\epsilon/T) + 1)$.

\section{Charge collective modes} \label{sec:modes}

\begin{figure*}
\begin{center}
\includegraphics[clip=true, width=1\textwidth]{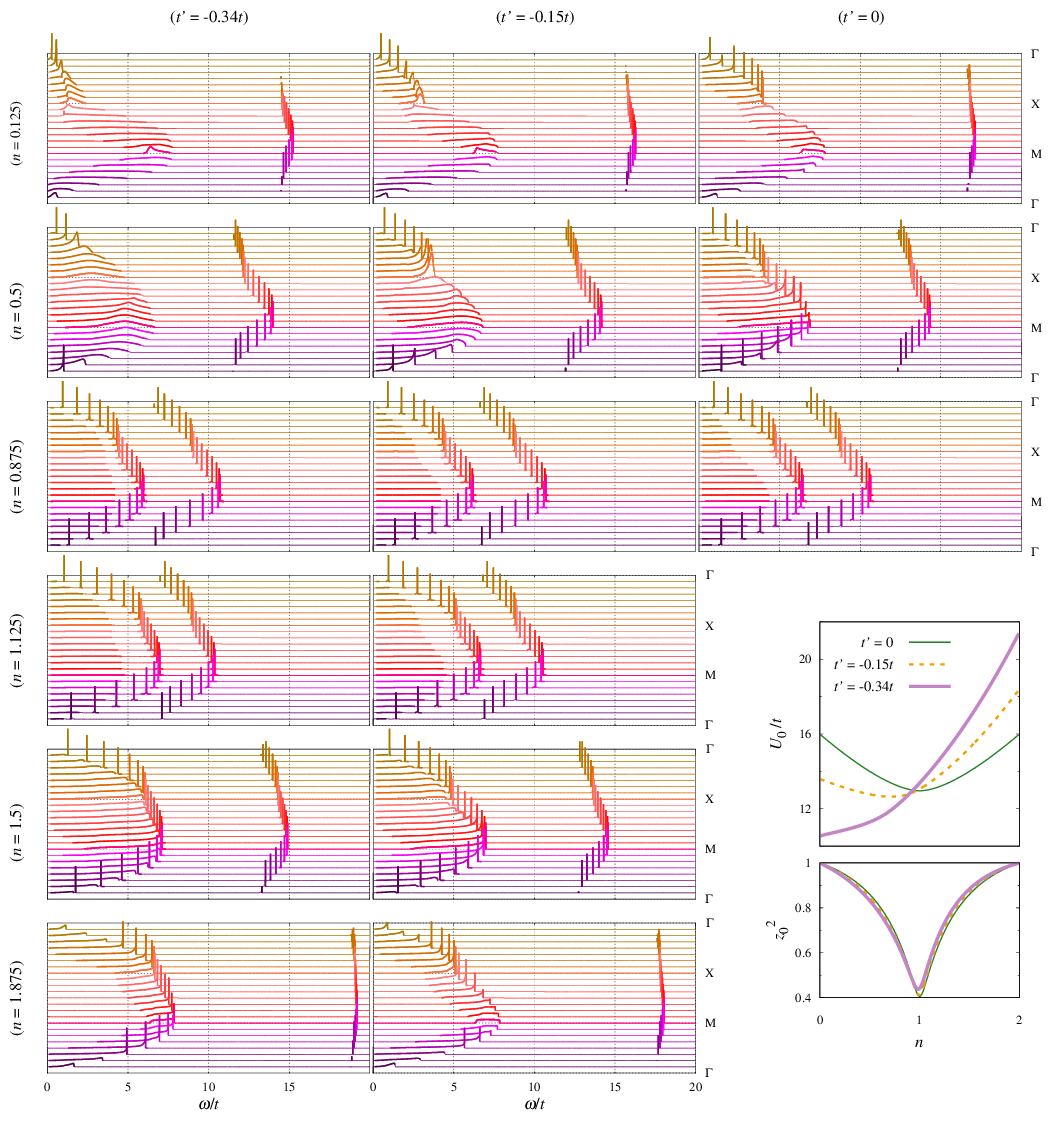} 
	\caption{(Color online) The imaginary part of the charge response function 
	$\chi_c(k)$, the coupling scale $U_0$, and the mass renormalization factor $z_0^2$ 
	for different values of NNN hopping $t'$ and density $n$. 
	The spectrum of ${\rm Im}\chi_c(k)$ is plotted for momenta along the path 
	linking ${\rm \Gamma}=(0,0)$, ${\rm X}=(\pi,0)$ and ${\rm M}=(\pi,\pi)$. 
	Parameters: $U= 10t$, $T=t/100$.}
	\label{fig:chi_c-U_10}
\end{center}
\end{figure*}

The lengthy but straightforward expansion of the terms~(\ref{eq:coeff}) shows that
they are invariant under the reversal of the doping sign when $t'=0$, so that 
$\chi_c(k)$ is symmetric~\cite{Dao17b}. This result stems from the particle-hole 
symmetry of the Hubbard model on a bipartite lattice, which pervades the dispersion 
and the susceptibility of the quasiparticles, as well as the paramagnetic 
saddle-point solution of the boson fields. 
However the symmetry does not hold for a finite $t'$.
As an illustration, the inelastic charge response given by ${\rm Im} \chi_c(k)$
is plotted for $t' = 0$, $-0.15t$, and $-0.34t$ at different 
densities in Fig.~\ref{fig:chi_c-U_10}. Since for $t'=0$ the charge susceptibility
is symmetric, it is shown only at densities $n<1$.

The spectra are composed of a broad continuum generated by incoherent single-particle-hole 
excitations, and the peaks of two collective modes above it. Both collective excitations
 have their minimum at $\Gamma$ and their maximum at M.
For the large coupling value $U=10t$, the correlation effects are important around 
half-filling where the continuum width is scaled down by the factor $z_0^2$ and a large 
portion of its intensity is transferred to the peaks. 
This prediction compares favorably with QMC simulations~\cite{Kun15}, within 
 the available energy resolution. Due to the numerical approximations inherent 
to the method and the high temperature ($T=t/3$) at which simulations are performed, the 
QMC spectra are Gaussian-like distributions. At low doping ($|\delta|\approx 0.05$) 
nearly all the intensity of the QMC signal is located at high energy, with the maximum 
between the positions of the two collective modes found within the slave-boson approach. 
With increasing the doping the bell of the QMC spectrum moves at a lower energy, although 
it retains a high-energy tail. In the slave-boson response, this corresponds to an 
increase of the continuum intensity, which also merges with the lower-energy collective 
mode. At the same time, the higher-energy peak goes up while its weight decreases. 
However, we shall note that the paramagnetic slave-boson result does not match the QMC
response at half-filling. This should not be surprising since the QMC ground-state 
is antiferromagnetic at zero doping~\cite{Hua17}.

The NNN hopping term visibly modifies the shape of the continuum since the 
bare quasiparticle dispersion $t_{\bf k}$ is changed. However $t'$ only has a small 
influence on the renormalization factor $z_0^2$, as shown in the lower right corner of 
Fig.~\ref{fig:chi_c-U_10}. Increasing $|t'|$ slightly narrows 
down (widens) the continuum for hole (particle) doping. The variation is a little 
more noticeable in the vicinity of $n=1$ where the width change is more important. 
This is in agreement with exact diagonalization results~\cite{Jia12,Wan14}.
In contrast, as discussed below in detail, the effect of varying $t'$ 
is more visible on the dispersion of the collective modes.

The higher-energy mode follows from the upper-Hubbard-band (UHB) with its excitation 
energy given in the strong coupling limit ($U \gg U_0$) by
$  \omega_{\rm UHB}({\bf k}) \approx U \sqrt{1 - \frac{U_0}{2U} \left( 1-3|\delta| 
 + (1-|\delta|) \frac{\varepsilon_{\bf k}}{\varepsilon_0}\right)}$, 
 which increases as the Coulomb coupling $U$~\cite{Dao17}. The coupling scale
\begin{equation}\label{eq:U_0}
 U_0 = - \frac{8 \varepsilon_0}{1 - \delta^2} 
\end{equation}
is plotted in Fig.~\ref{fig:chi_c-U_10}, and
\begin{equation} \label{eq:epsk}
 \varepsilon_{\bf k} = \frac{2}{L} \sum_{{\bf q}} t_{{\bf q}+{\bf k}} n_F(E_{\bf q}). 
\end{equation} 
In this regime the mode starts around the energy $U + U_0 (|\delta|-1/2)$ and extends 
over a range $\approx (1 - |\delta|)U_0/2$. In the opposite limit ($U \ll U_0$) the 
dispersion is 
$  \omega_{\rm UHB}({\bf k}) \approx \frac{U_0}{2} \sqrt{1 + \frac{U}{2U_0} 
 \left( 1+7\delta^2 - (1-\delta^2)\frac{\varepsilon_{\bf k}}{\varepsilon_0}\right)}. $
Although the weak-coupling expression is a rough approximation for the coupling value 
$U=10t$, it nevertheless yields the qualitative behavior of the mode. It locates 
the bottom of its dispersion around $U_0/2 + U \delta^2$, and gives a width of about 
$U(1-\delta^2)/4$. 
As illustrated in Fig.~\ref{fig:chi_c-U_10}, doping the system increases the 
collective-excitation energy and narrows down its dispersion. 
The comparison of the spectra at a fixed doping shows that $t'$
softens (hardens) the UHB mode for large positive (negative) values of doping. 
Indeed a negative $t'/t$ flattens the quasiparticle dispersion around $\Gamma$ 
while it increases their velocity around M (see Fig.~\ref{fig:dispersion}). 
As a result, the minimum of the average kinetic energy $\varepsilon_0$ is moved
from $n=1$ for $t'=0$ to $n>1$ for $t'/t<0$. This reduces $U_0$ at $n<1$ and 
increases it at $n>1$ (see Fig.~\ref{fig:chi_c-U_10}). 

\begin{figure}
  \includegraphics[clip=true, width=0.49\textwidth]{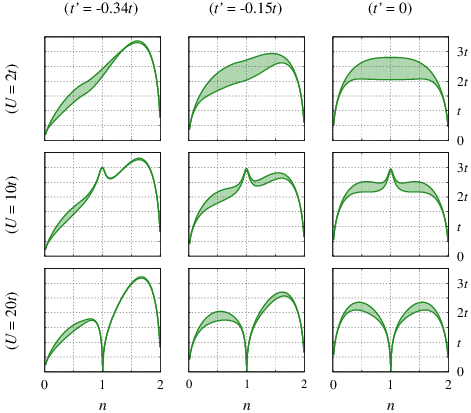}
  \caption{(Color online) Dispersion of the zero-sound velocity as a function of 
  the density, for different values of $t'$ and $U$. Parameter: $T=t/100$.}
  \label{fig:zsvelocity}
\end{figure}

The lower-energy collective excitation, called the ZS mode, is 
located between the upper edge of the continuum $\omega_{\rm cont}({\bf k})$ 
and the UHB mode. Contrary to the UHB mode, its energy vanishes at $\Gamma$
as its dispersion is linear at long wavelength. In this limit one then defines
the ZS velocity as $c_s({\bf \hat{k}}) = \omega_{\rm ZS}({\bf k})/|{\bf k}|$.
It is anisotropic on the square lattice, with the minimum in the X-direction and
the maximum in the M-direction. Fig.~\ref{fig:zsvelocity} shows the density
dependence of its dispersion for different values of coupling and NNN hopping 
amplitude. 
Strong correlations increase the ZS velocity around half filling for $U$ smaller 
than the critical value $U_c = 2(8/\pi)^2t \approx 12.97t$. Above the 
critical coupling, $c_s({\bf \hat{k}})$ vanishes at half filling as the state 
is insulating. 
For $t'=0$ the density dependence is symmetric from either side 
of $n=1$. Increasing the amplitude of $t'$ decreases (increases) the velocity 
at positive (negative) doping, in accord with the modification of
the bare quasiparticle dispersion $t_{\bf k}$. As previously noted for the Fermi 
velocity, the anisotropy of $c_s({\bf \hat{k}})$ is reduced at negative doping
by the NNN hopping.

\begin{figure}
  \includegraphics[clip=true, width=0.49\textwidth]{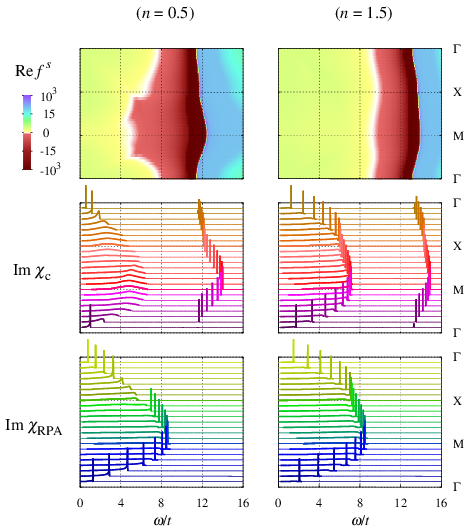}
  \caption{(Color online) Energy and momentum dependence of ${\rm Re} f^s(k)$,
  and comparison between the slave-boson charge response $\chi_c(k)$ and the RPA 
  result $\chi_{\rm RPA}(k)$ for density $n=0.5$ and 1.5. 
  Parameters: $t'=-0.34t$, $U=10t$, $T=t/100$.}
  \label{fig:fs}
\end{figure}

Fig.~\ref{fig:chi_c-U_10} shows that for $t'=0$ a large doping suppresses 
the ZS mode in a large part of the Brillouin zone. For a finite $t'$, a positive 
doping enhances the inhibition, while a negative doping 
favors the ZS mode. The change of the quasiparticle dispersion is not the 
sole cause of the ZS mode damping. There is also a dynamical screening of the 
interaction induced by correlations. In order to discuss this effect, we write 
the charge response function as
\begin{equation}
 \chi_c(k) = \frac{\chi_0(k)}{1+ f^s(k) \chi_0(k)} 
\end{equation}
with the Lindhard function
\begin{equation}
 \chi_0(k) = \frac{2}{L} \sum_{\bf q} \frac{n_F(E_{{\bf q} + {\bf k}}) - n_F(E_{\bf q})} 
 {(\omega + \ii 0^+) - (E_{{\bf q} + {\bf k}} - E_{\bf q})} .
\end{equation}
Here $f^s(k) = \chi_c(k)^{-1} - \chi_0(k)^{-1}$ represents an effective interaction
that reduces to $U/2$ in the weak-coupling limit. There, the 
random-phase approximation (RPA) result
\begin{equation}
 \chi_{\rm RPA}(k) = \frac{\chi_0^{(0)}(k)}{1+\frac{U}{2} \chi_0^{(0)}(k)}
\end{equation}
is recovered~\cite{Dao17}. 
Since $\chi_0({\bf k})$ is real above $\omega_{\rm cont}({\bf k})$ and has a 
negative value, this explains how the pole associated to the ZS mode appears 
just beyond the continuum upper edge when increasing $U$. However, as one can 
expect, the RPA perturbation approach breaks down at large coupling. 
Fig.~\ref{fig:fs} shows the strong dependence on momentum and frequency of the 
complex function $f^s(k)$ for $U=10t$. Its value, indeed, can drastically differ 
from $U/2$. It even goes to infinity at an energy $\omega \sim U$, which gives 
rise to the pole of the UHB mode. 
The RPA response neither possesses the higher-energy mode, nor accounts for the 
renormalization of the continuum width. Going back to the causes of the ZS mode 
suppression, one can note that $\chi_{\rm RPA}(k)$ has no peak around X at 
density $n=0.5$, contrary to $n=1.5$. The damping here is only ascribed to the 
differences in the quasiparticle dispersion. However in the slave-boson response 
at $n=0.5$, the suppression extends up to a larger region of the ${\bf k}$-space. 
The enhancement of the damping stems from the screening of the effective potential 
$f^s(k)$ which even turns it negative around M. In contrast, for density $n=1.5$, 
the screening is limited to higher energies and does not prevent the ZS mode.
As recently revealed by the detailed comparison~\cite{Kun15} between QMC 
simulations and the RPA, strong correlations persist in multiparticle responses
over doping values larger than the one for single-particle properties, especially
in the charge channel. In agreement with the QMC results, the slave-boson charge
response at low density $n\lesssim 0.5$ is essentially formed by a broad 
featureless continuum in a large part of the Brillouin zone, in contrast to 
the RPA intensity which shows a peaked maximum at the continuum upper edge. 
The latter results from a resonance with the ZS mode, and forms a well separated
peak at low temperature. The RPA pole is present around ${\bf k}={\rm M}$ even 
at density as low as $n=0.1$, while it is suppressed by correlations in the 
slave-boson response.

\section{Conclusion} \label{sec:conclusion}

We have calculated the charge excitation spectra of the $t-t'-U$ Hubbard
model on the square lattice in the thermodynamical limit
within KR slave boson formalism. While our approach reduces to RPA in the weak
coupling regime, and while both approaches may be used
in the thermodynamical limit to resolve the full momentum dependence of the
spectra, they exhibit significant differences from the intermediate coupling
regime on. In particular, the former is sensitive to the renormalization of
the continuum width and displays a higher-energy mode originating from the
upper Hubbard band. Furthermore sharp peaks artificially predicted by the RPA
that are absent in QMC simulations are fully damped in our approach, too.
In fact, a good agreement between our paramagnetic calculations and QMC
simulations has been found, provided the latter are, of course, performed
outside a magnetic phase.  
The influence of $t'$ on the shape of the response continuum is strongest for 
large doping, though the correlation-induced renormalization of its width is 
barely affected. Increasing $-t'$ softens (hardens) the high-frequency UHB
collective mode at positive (negative) doping. When approaching half-filling,
the mode loses most of its $t'$ dependence while it gets broader, and is
shifted to lower frequency. 
For low (high) density $n \lesssim 0.5$ ($n \gtrsim 1.5$), 
the suppression of the ZS collective mode is favored (reduced) by $t'$.
Its damping is enhanced by a dynamical screening of the interaction 
most effective for large momenta. The averaged ZS velocity 
decreases (increases) with increasing $-t'$ for hole (particle) doping, 
following the change in the quasiparticle dispersion. 
The above discussed collective modes should be observable in the
paramagnetic phase of the superconducting cuprates by means of inelastic
neutron scattering experiments. This, in fact, could even be a way to
measure $t'$.

\acknowledgments

We gratefully thank T. Kopp for several stimulating discussions. 
The authors acknowledge the financial support of the French Agence Nationale de la Recherche (ANR), through the program Investissements d'Avenir (ANR-10-LABX-09-01), LabEx EMC3.

\end{document}